\documentclass[twocolumn,preprintnumbers,amsmath,amssymb,pra]{revtex4}

\usepackage[colorlinks,linkcolor=blue,anchorcolor=blue,citecolor=blue]{hyperref}
\usepackage{txfonts}
\usepackage{graphicx,hyperref}
\usepackage{subfigure}
\usepackage{dcolumn}
\usepackage{bm}
\usepackage{braket}
\usepackage{upgreek}
\usepackage{extarrows}
\usepackage{setspace}
\usepackage{float}
\usepackage{makecell}
\usepackage{rotating} 

\begin{document}
	\title{Photon-assisted tunneling resonantly controlling spin current of a spin-orbit-coupled atom	 in a toroidal trap}
	\author{Zhiqiang Li$^{1}$}
	\author{Xiaoxiao Hu$^{1}$}
	\author{Zhao-Yun Zeng$^{2}$}
	\author{Ai-Xi Chen$^{1}$}
	\author{Xiaobing Luo$^{1,2}$}
	\altaffiliation{Corresponding author: xiaobingluo2013@aliyun.com}
	\affiliation{$^{1}$Department of Physics, Zhejiang Sci-Tech University, Hangzhou, 310018, China}
	\affiliation{$^{2}$School of Mathematics and Physics, Jinggangshan University, Ji'an 343009, China}
	\date{\today}
	\begin{abstract}
		The periodic flashing potential has proven to be a powerful tool for investigating directed atomic currents. By applying the flashing ring-shaped potential to spin-orbit (SO) coupled, noninteracting Bose-Einstein condensate (BEC) systems, through photon-assisted tunneling (resonance) techniques, we demonstrate the generation of tunable alternating (AC) spin and atomic mass currents that can be precisely controlled in terms of direction and strength. The underlying mechanism behind this phenomenon is that the flashing potential supplies enough photons to induce Rabi oscillations and provides momentum transfer for spin and atomic transport. As the single-particle ground state of the unperturbed SO-coupled BEC depends on the Raman coupling strength, we demonstrate how to generate and control AC spin currents in the cases where the initial state resides in a single-well or double-well phase. In particular, we realize and explain the mechanism of generating a net AC spin current without mass current through single-photon resonance processes. It is shown that these interesting resonance phenomena can be analytically described only by the simple three-level model, which creates the possibility of transparent controls of spin dynamics.
	\end{abstract}$ $
	\maketitle

	\section{Introduction}
	Spin-orbit coupling (SOC) is a significant phenomenon in quantum mechanics that links the spin of an electron to its orbital motion, and it has various concrete manifestations in practical applications, such as the fine structure in atomic spectra, spintronics, topological insulators, topological semimetals, and quantum computing technologies. In atomic experiments, SOC has been successfully implemented experimentally in both bosonic and fermionic atoms, by using a pair of counter-propagating Raman laser beams that couple two hyperfine states of atoms \cite{Y.-J. Lin,Pengjun Wang,Lawrence W. Cheuk,L. Huang,Z. Wu,Z.-Y. Wang}. The realization of SOC in ultracold atoms creates a highly tunable experimental platform, which is particularly useful for quantum simulation and enhances our understanding of multi-component Bose-Einstein condensate (BEC). Specifically, BEC with SOC exhibit  a rich variety of ground-state phases, including stripe, plane-wave, and zero-momentum phases \cite{C. Wang,T.-L. Ho,S. Sinha,Y. Li,Y. Deng,H. Zhai}. The existence of these phases provides abundant material for both theoretical and experimental investigation. In addition, theoretical studies and experiments have explored the nonequilibrium spin dynamics in systems with Raman-induced SOC \cite{C. Qu,M. C. Beeler,Qizhong Zhu,C.-H. Li}. For instance, by quenching a spin-orbit (SO) coupled BEC, the Zitterbewegung oscillations for the center-of-mass motion have been observed \cite{C. Qu}.
		
	On the other hand, spin current, which refers to the flow of spin, could potentially replace charge currents to address the issue of waste energy due to Joule heating \cite{F. Meier}. This phenomenon is a  key ingredient in applications such as spintronics \cite{J. Fabian} and has been generated in semiconductors \cite{C. Leyder,A.A. High} and magnetic insulators \cite{L.J.Cornelissen,D. A. Bozhko}. In the realm of ultracold atomic gases, various experimental schemes have been exploited to realize spin transport \cite{M. C. Beeler,Y. Eto,Qizhong Zhu,K. Nakata,C.-H. Li,B. Divinskiy,H. J. Lewandowski,X. Du,X. Du2,L. Tian,Y. Sekino}. For instance, the spin Hall effect has been observed through the use of a meticulously designed spin- and space-dependent vector potential \cite{M. C. Beeler}. Moreover, spin current has been generated and observed through quench dynamics in a SO-coupled BEC \cite{C.-H. Li} and anomalous spin segregation has been detected in both Bose and Fermi gases \cite{H. J. Lewandowski,X. Du}. Furthermore, by leveraging dynamical decoupling techniques, it has been possible to generate longer spin-coherent currents in a Bose gas by periodically applying $\pi$ pulses \cite{Y. Eto}. 
	
	Persistent current in closed circuits for ultracold atoms enables fundamental studies of superfluidity \cite{S. Gupta,L. Corman,G. Del Pace}. Annular geometries have played a significant role in studying such persistent currents, and extensive work has been conducted to investigate the flow of ultracold atoms in ring-shaped geometries for both bosonic and fermionic gases \cite{S. Gupta,L. Corman,G. Del Pace,T. Bland,S. Simjanovski,F. Perciavalle,N. Pradhan,K. Xhani,N. Grani,K. Mukherjee,D. Poletti,C. E. Creffield,M. Heimsoth,Zhiqiang Li}. While there has been progress in generating atomic currents using flashing ratchets in single-component systems \cite{D. Poletti,C. E. Creffield,J. Gong,DenisovS,S. Denisov,Sergey Denisov,Tobias Salger,M. Heimsoth,Emil Lundh,Mark Sadgrove,Anatole Kenfack,D. H. White,Jiating Ni,Clement Hainaut,Dario Poletti,Wen-Lei Zhao,T. S. Monteiro,L Morales-Molina and S Flach,Stefano Longhi,Wen-Lei Zhao2,J.-Z. Li,Zhiqiang Li,Peter Hanggi}, the application of these coherent engineering schemes to achieve spin transport, particularly in SO-coupled BEC systems, represents an area of research that has not yet been investigated. In this paper, we are specifically interested in uncovering  the coherent spin current in a noninteracting BEC (or a SO-coupled atom) confined within a toroidal trap, using a flashing potential. By utilizing photon-assisted tunneling (PAT)—a powerful tool for controlling quantum tunneling \cite{M. Grifoni,C. Sias,C. Weiss,N. Teichmann,Q. Xie,R. Ma,A. Eckardt,L. Morales-Molina} and transport processes \cite{P. K. Tien}—we can achieve a net alternating current (AC) spin current that can be precisely controlled in both direction and amplitude. We further demonstrate that the spin and mass currents exhibit distinct dynamical behaviors in response to the different phases of the single-particle energy dispersions in SO-coupled BEC. Moreover, we derive analytical results for the generation of spin and mass currents using a simple three-level model, rendering the control strategies more transparent and effective.
	
	The structure of this paper is as follows. In Section \ref{II}, we introduce our model, which includes two primary components: a noninteracting BEC with SOC and an external periodically flashing ring-shaped potential. We discuss the single-particle dispersion relation of the SO-coupled BEC. In Section \ref{III}, we introduce the resonance engineering scheme within the extended Hilbert space. Sections \ref{IV} and \ref{V} present our findings on the generation and manipulation of spin current and mass current in the single-well and double-well phases, respectively. We conclude with a summary of our results in Section \ref{VI}. Additional technical details are provided in Appendices \ref{Appendix A}-\ref{Appendix B}.

	\section{Model and dispersion} \label{II}
	We consider a one-dimensional system of pseudo-spin-1/2 noninteracting BEC with Raman-induced SOC, confined in a toroidal trap where the radius $R$ of the trap is much greater than its thickness $r$ (i.e., $R \gg r$). Such a SOC via Raman coupling has been first experimentally realized in \cite{Y.-J. Lin}. The model Hamiltonian is given by the dimensionless linear Gross-Pitaveskii (GP) equation (taking $\hbar=m=1$),
	\begin{equation}\label{eq1}
		i\frac{\partial}{\partial t}\ket{\psi(t)}=\hat{H}(t)\ket{\psi(t)},
	\end{equation}
	where \(\hat{ H}(t) = \hat{H}_0 + V(x,t) \). The condensate is subjected to an external potential that varies periodically in time and has an average value of zero, denoted as 
	\begin{equation}\label{eq2}
	V(x,t)=K\sin{(\kappa x)}\sin{(\omega t)},
	\end{equation}
	where $ x $ is the coordinate subject to the periodic boundary condition $ 0 \leq x < 2\pi $, $ K $ and $ \omega $ represent the amplitude and angular frequency of the flashing potential, respectively, and $ \kappa $ is the wave vector of the spatial periodic potential, which is an integer in our study. The single-particle Hamiltonian  is defined as 
	\begin{equation}\label{eq3}
	\hat{H}_0=\begin{pmatrix}
	\frac{(\hat{p}-k_0)^2}{2}+\frac{\delta}{2}& \frac{\Omega}{2}\\
	\frac{\Omega}{2}& \frac{(\hat{p}+k_0)^2}{2}-\frac{\delta}{2}
	\end{pmatrix},
	\end{equation}
	where $ \hat{p} = -i \frac{\partial}{\partial x} $ is the momentum operator in the longitudinal direction, $\Omega$ and $\delta$ signify the  Raman coupling strength and detuning, respectively, and $k_0$ represents the wave number of the Raman laser that couples the two spin (hyperfine) states. For simplicity, we confine our analysis to the case of zero detuning, $\delta=0$.  Owing to the periodic boundary condition of the toroidal trap, the angular momentum $ \hat{p} $ is quantized, with $\hat{p}\ket{n}=n\ket{n}$, where $n$ is an integer and $\braket{x|n}=1/\sqrt{2\pi}e^{inx}$. Consequently, the quantum states can be expanded in terms of the plane wave basis, 
	\begin{equation}\label{eq4}
	\ket{\psi(t)}=\sum_{n=-\infty}^{\infty}\frac{1}{\sqrt{2\pi}}e^{inx}
	\begin{pmatrix}
		\phi_\uparrow (t) \\ \phi_\downarrow (t)
	\end{pmatrix}. 
	\end{equation}
	
\begin{figure}[htp]	
		\centering
		\includegraphics[width=11.5cm]{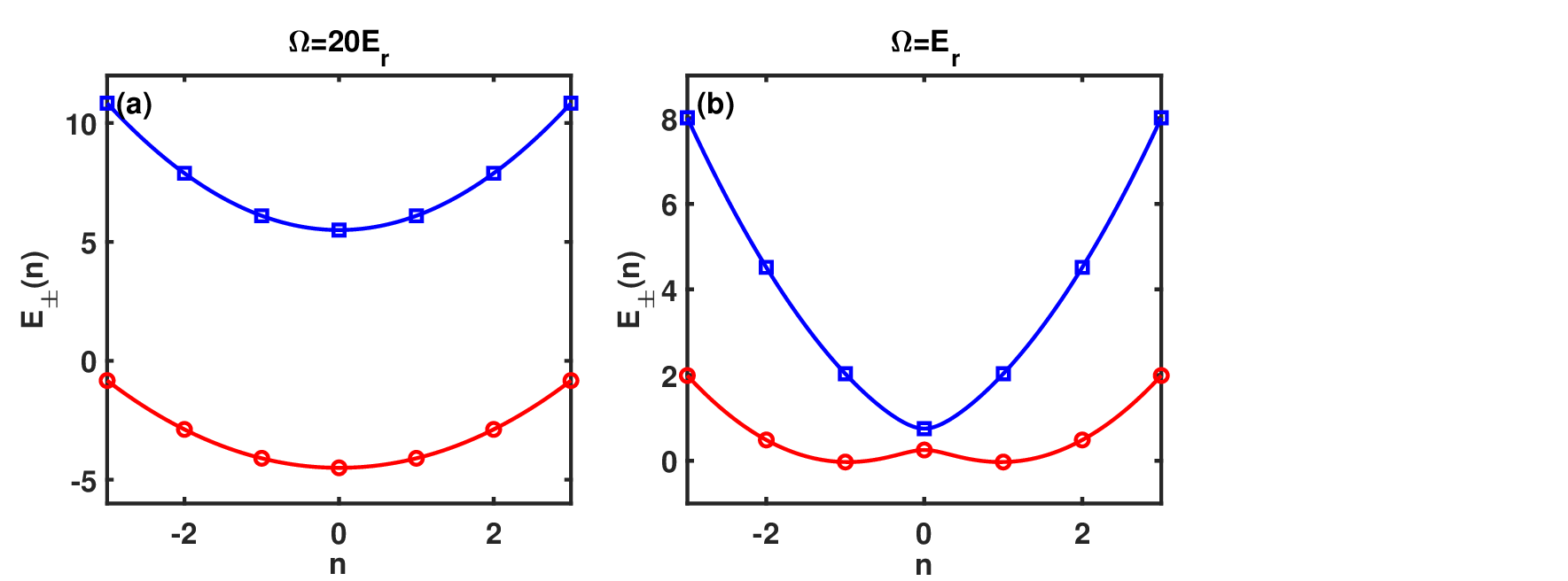}
		\caption{Single-particle dispersion $E_\pm(n)$ for a Raman-induced SO-coupled BEC with different coupling strengths \(\Omega\), (a): in the single-well phase with \(\Omega = 20E_r\) and (b): in the double-well phase with \(\Omega = E_r\). Due to the periodic boundary condition, the quasi-momentum \( n \) (expressed in units of \( k_0 \)) is quantized and can only assume integer values.} \label{fig1}
	\end{figure}

	The unperturbed single-particle Hamiltonian $ H_0 $ featuring this type of Raman-induced SO coupling can be written in the form of a momentum-dependent Zeeman field, 
	\begin{equation}\label{eq5}
		\hat{H}_0=\frac{\hat{p}^2}{2}+E_r-k_0\hat{p}\sigma_z+\frac{\Omega}{2}\sigma_x,
	\end{equation}
	where $E_r=\frac{k_0^2}{2}$ is the recoil energy. The single particle dispersion is given by
	\begin{equation}\label{eq6}
		E_\pm(n)=\frac{n^2}{2}+E_r\pm\sqrt{(nk_0)^2+\frac{\Omega^2}{4}},
	\end{equation}
	which features two distinct branches, see Fig. \ref{fig1}.
	The corresponding eigenstates read
	\begin{equation}\label{eq7}
		\ket{n+}=\frac{1}{\sqrt{2\pi}}e^{inx}	
		\begin{pmatrix}
			\sin\theta \\ \cos\theta
		\end{pmatrix},\quad
		\ket{n-}=\frac{1}{\sqrt{2\pi}}e^{inx}	
		\begin{pmatrix}
			\cos\theta \\ -\sin\theta
		\end{pmatrix},
	\end{equation}
	 with the variational parameter
	\begin{equation}\label{eq8}
	\theta(n)=\arcsin\bigg[\frac{1}{2}\bigg(1-\frac{nk_0}{\sqrt{(nk_0)^2+\Omega^2/4}}\bigg)\bigg]^\frac{1}{2}.
	\end{equation}
	For $\Omega<4E_r$, the ground band $E_-(n)$  displays a double-well structure, which we term the double-well phase, with two degenerate minima appearing near
	\begin{equation}\label{eq9}
		\bar{n}=\pm k_0\sqrt{1-\bigg(\frac{\Omega}{4E_r}\bigg)^2}.
	\end{equation}
	As \(\Omega\) increases beyond \(\Omega_c = 4E_r\), the double minima merge into a single minimum at \(n=0\),  displaying the single-well structure. 
	
	Due to the different phases of the ground states, the system yields rather different results for the dynamics of spin current and mass current. As we will illustrate, when starting from the ground state of \(\ket{0-}\) in the single-well phase, we can harness quantum resonance to produce a purely alternating spin current with no associated mass current. In the double-well phase, when starting from the superposition of two degenerate states with minimum energy, we note that the time-averaged spin current, induced by quantum resonance, is independent of the difference in the superposition coefficients, whereas the time-averaged mass current is linearly proportional to the difference in these coefficients. 

	\section{Resonance current} \label{III} 
	To begin, we introduce the canonical momentum for the $\sigma$ component (either $\uparrow$ or $\downarrow$), denoted as $p_\sigma(t)$, for the (pseudo-) spinor $\ket{\psi(t)} = [\psi_\uparrow(x,t), \psi_\downarrow(x,t)]^T$ (where $T$ denotes transposition),
	\begin{equation}\label{eq11}
		p_\sigma(t)=\int_0^{2\pi}dx\psi^*_\sigma(x,t)\hat{p}\psi_\sigma(x,t),
	\end{equation}
	as well as its long-time average,
	\begin{equation}\label{eq10}
		\bar{p}_\sigma=\lim_{t\to \infty}\frac{1}{t}\int_{t_0=0}^tdt^{\prime}p_\sigma(t^\prime),
	\end{equation}
	where $\psi_\sigma(x,t)$ represents the two pseudospin components  of the quantum state $\ket{\psi(t)}$. From now on, in all formulas, the overline denotes the time average. 
	
	In the context of BEC with Raman-induced SOC, the spin current represents the relative current density between the two spin components and is phenomenologically defined as $I_s=I_\uparrow-I_\downarrow$ \cite{M. C. Beeler,C.-H. Li}, where $I_\uparrow$ and $I_\downarrow$ are the macroscopic current densities for spin-up and spin-down, respectively. Here, the particle current $ I_\sigma(t) $ carrying an upward (downward) spin is defined as $I_\sigma(t)=\int_0^{2\pi}dx\psi^*_\sigma(x,t)\Big(\hat{p}\pm k_0\Big)\psi_\sigma(x,t)$ within the framework of a Raman-induced SO-coupled BEC ($-$ for $\uparrow$ and $+$ for $\downarrow$). The formal relationship between the spin current and the canonical momentum $ p_\sigma $ is presented, along with its corresponding time-averaged formula, which is detailed in Appendix \ref{Appendix A},
	\begin{equation}\label{eq12}
	 	I_s(t)=p_\uparrow(t)-p_\downarrow(t)-k_0,\quad\bar{I}_s=\lim_{t\to\infty}\frac{1}{t}\int_{0}^t dt^\prime I_s(t^\prime).
	\end{equation}
    By contrast, the (atomic) mass current is described as $I_m = I_\uparrow + I_\downarrow$. Accordingly, we derive expressions for the mass current and its long-time average from the canonical momentum, 
    \begin{equation}\label{eq13}
    	I_m(t)=p_\uparrow(t)+p_\downarrow(t)-k_0\Delta N(t),\quad\bar{I}_m=\lim_{t\to\infty}\frac{1}{t}\int_{0}^t dt^\prime I_m(t^\prime),
    \end{equation}
	where $\Delta N(t)=\int_0^{2\pi}dx\Big(|\psi_{\uparrow}(x,t)|^2-|\psi_{\downarrow}(x,t)|^2\Big)$ represents the population imbalance between the two components. The definition indicates that the spin current is shifted by $-k_0$ relative to the difference in the canonical momentum of the two components, $ p_\uparrow(t) - p_\downarrow(t) $. Interestingly, we observe that when the time-averaged population imbalance is zero, the dynamics of the mass current may be distinct from those of the sum of the canonical momenta $p_\uparrow + p_\downarrow$ over time, yet the time-averaged mass current remains equal to the time-average of $p_\uparrow + p_\downarrow$.
	
\begin{figure}[htp]	
		\centering
		\subfigure{
			\centering
			\includegraphics[width=9cm]{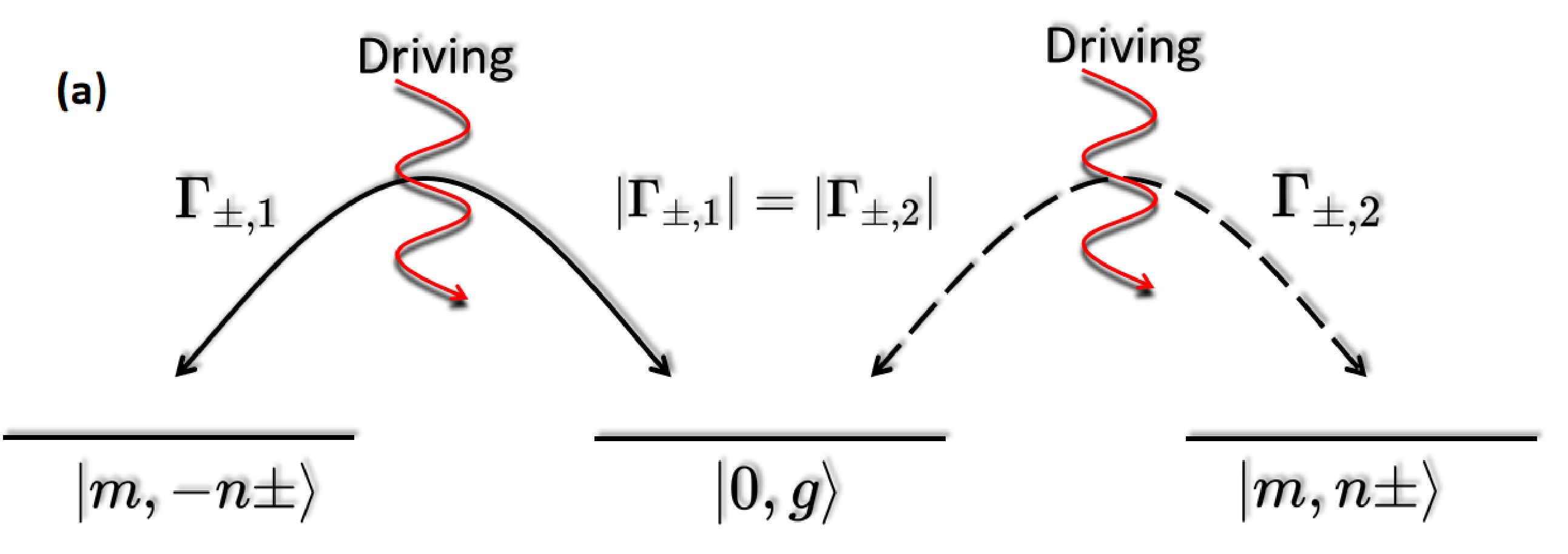}
			\label{fig2a}
		}\\
		\subfigure{
		\centering
		\includegraphics[scale=0.277]{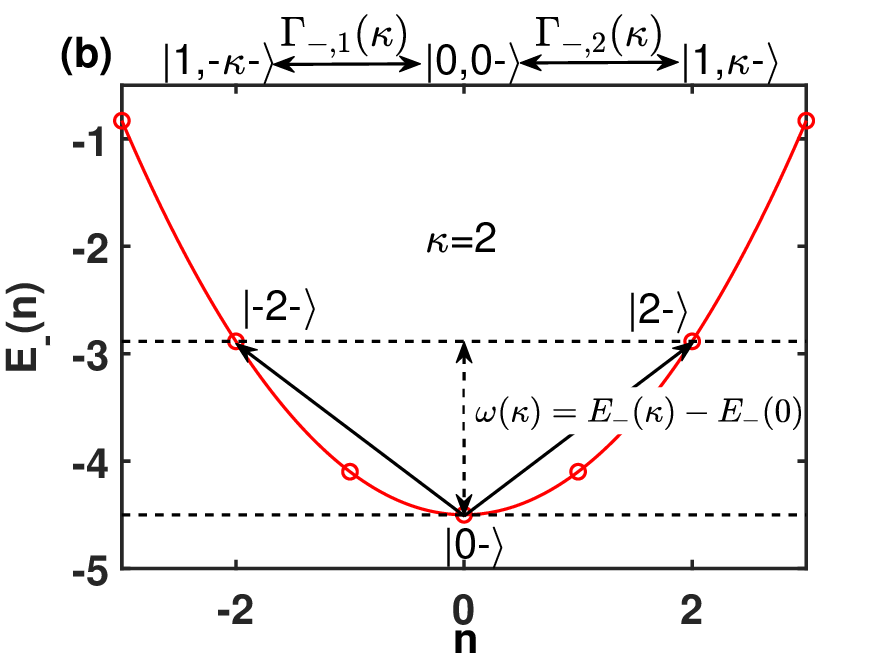}
		\label{fig2b}
		}\
		\subfigure{
		\centering
		\includegraphics[scale=0.277]{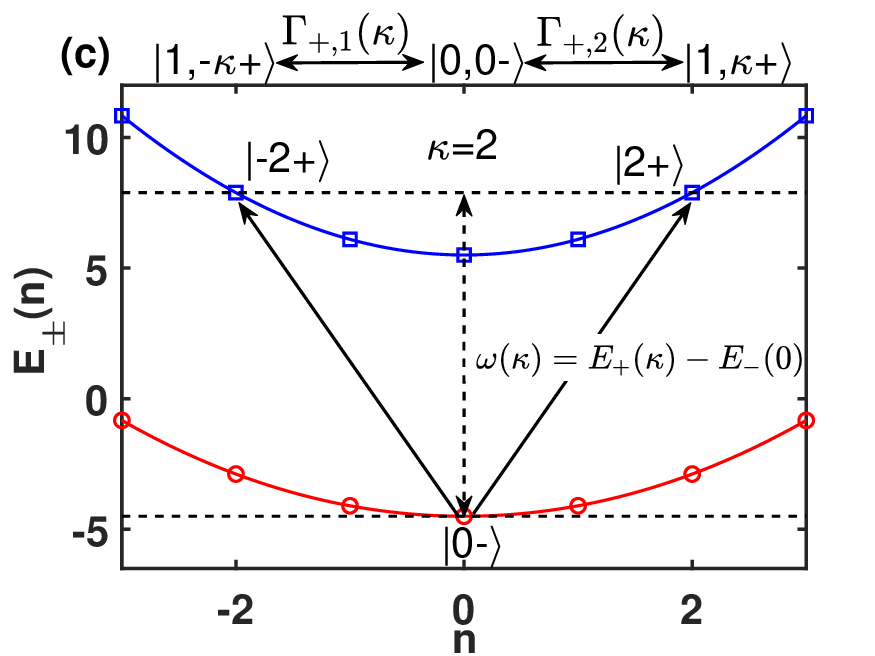}
		\label{fig2c}
		}\
		\caption{Resonance diagram: (a) The initial state \(\ket{0,g}\) is mixed with the excited states \(\ket{m,\pm n\pm}\) for $m$-photon resonance, with the effective coupling strengths \(\Gamma_{\pm,1}\) between \(\ket{0,g}\) and \(\ket{m,- n\pm}\), and \(\Gamma_{\pm,2}\) between \(\ket{0,g}\) and \(\ket{m,+ n\pm}\). \( |\Gamma_{\pm,1}| = |\Gamma_{\pm,2}| \). Here, the subscript “\(\pm\)" denotes the mixing of the initial state to the resonant states in the upper (lower) dispersion branch, respectively. (b) and (c) depict the schematic representation of the first-photon ($m=1$) resonance process in the single-well phase: (b) shows the initial state \(\ket{0-}\) resonantly coupling to excited states \(\ket{\pm\kappa-}\) in the lower branch with a resonance frequency \(\omega(\kappa)=E_-(\kappa)-E_-(0)\), and (c) shows the initial state \(\ket{0-}\) resonantly coupling to excited states \(\ket{\pm\kappa+}\) in the upper branch with a resonance frequency \(\omega(\kappa)=E_+(\kappa)-E_-(0)\), respectively. Here we choose the wave vector of the spatial potential \(\kappa=2\) as an example, which provides the momentum transfer between the initial state \(\ket{0-}\) with the excited states \(\ket{\pm2-}\) and \(\ket{\pm2+}\).}\label{fig2}
	\end{figure}

	Subsequently, we delve into the mechanisms underlying the generation and control of spin and mass currents through the utilization of photon-assisted tunneling (resonance). We are particularly fascinated by a distinctive dynamical phenomenon in the Floquet SO-coupled system when the conditions for quantum resonance are satisfied, specifically \( E_{\pm}(n) - E_g = m\omega \), where \( E_{\pm}(n) - E_g \) denotes the unperturbed energy level difference between the excited states \(\ket{n\pm}\)  and the ground state \(\ket{g}\).
	Particularly, the nature of the ground state is contingent upon the dispersion, as illustrated in Fig. \ref{fig1}. For instance, in cases where the ground band of the energy-quasimomentum dispersion assumes a single-well structure (single-well phase), the ground state \( |g\rangle \) corresponds to \( |0,-\rangle \). We anticipate that the spin and mass currents will manifest distinct dynamical behaviors in response to different dispersions. The resonance condition tells us that the ground states absorb the energy of $m$ photons to match the energy gap between the excited and ground states, thereby facilitating a quantum transition. To give an excellent account of the resonant dynamics, we follow the perturbative method of \cite{M. Heimsoth,I. Shavitt,J. Hausinger} and generalize it to the SO-coupled system. Our starting point for the study is  the extended Hilbert space, which is spanned by the unperturbed Floquet states $\ket{m,n\pm}=\frac{1}{\sqrt{2\pi T}}e^{-im\omega t}\ket{n\pm}$, where $T=\frac{2\pi}{\omega}$ is the period of driving. In the extended Hilbert space, the system's dynamics are governed by the quasienergy operator $\hat{Q}=\hat{H}(t)-i\frac{\partial}{\partial t}$. The matrix representation of $\hat{Q}$ in the basis of states $\ket{m,n\pm}$ can be expressed as
	\begin{equation}
		Q=\left(\begin{array}{ccccc}
			\quad\ddots &\vdots &\vdots &\vdots &\begin{sideways}$\ddots$\end{sideways} \\
			\cdots &H_{\mathrm{0}}+\omega & H_{-1} & H_{-2} & \cdots\\
			\cdots	& H_{1} & H_{\mathrm{0}} & H_{-1} &\cdots \\
			\cdots	& H_2& H_{1} & H_{\mathrm{0}}-\omega &\cdots \\
			\begin{sideways}$\ddots$\end{sideways}&\vdots &\vdots &\vdots &\quad\ddots
		\end{array}\right),\label{eq15}
	\end{equation}
	where the matrix elements of the block matrix $H_m$ are given by $\bra {n\pm}\hat{H}_m\ket{n'\pm}$, with $\hat{H}_m$ representing the Fourier series expansions of the Hamiltonian, i.e. $\hat{H}(t)=\sum_m \hat{H}_me^{-im\omega t}$.  Diagonalizing the $Q$ matrix allows us to determine the quasienergies of the Floquet system. 
	When the initial state is prepared as the ground state \(\ket{0,g}\) of the unperturbed Hamiltonian, we may expect that the driving will mix the initial state with other unperturbed  Floquet states \(\ket{m,n\pm}\) that satisfy the resonance condition \(E_\pm(n) - E_g = m\omega\). The $Q$ matrix can subsequently be truncated to an effective three-level model that includes the relevant resonant states.
	This process can be understood as a periodic oscillation between the initial state and the excited states, as illustrated in Fig. \ref{fig2} (a). Specifically, the initial state \(\ket{g}\) absorbs $m$ photons to transition to the excited states \(\ket{-n\pm}\) and \(\ket{n\pm}\), and the excited states emit $m$ photons to return to the initial state, indicating the mixing between \(\ket{0,g}\) and states \(\ket{m,-n\pm}\) (\(\ket{m,n\pm}\)) in the extended Hilbert space. The momentum $\pm n$ of the excited states is provided by the wave numbers of the periodic spatial potential. In this paper, we focus solely on the first-order dynamics (the first photon resonance) for simplicity, which corresponds to the case \(m = 1\).

	\section{Single well phase} \label{IV}
	We begin by examining the scenario where the dispersion of the ground (lower) band exhibits a single minimum at \( n = 0 \),  as illustrated in Fig. \ref{fig1} (a). We examine the first photon resonance ($m=1$), with the resonance condition \( E_\pm(\kappa) - E_-(0) = \omega \), where the energy gap between the excited states $\ket{\pm\kappa-}$ $(\ket{\pm\kappa+})$ in the lower (upper) branch of the energy dispersion and the ground state \( |0-\rangle \) is bridged by the energy of a single photon. In this scenario, the dynamics are confined to the truncated extended Hilbert space spanned by the three resonant states \(\{\ket{1,-\kappa\pm},\ket{0,0-},\ket{1,\kappa\pm}\}\), as illustrated in Figs. \ref{fig2} (b) and (c). By applying the time-independent Van Vleck degenerated perturbation theory \cite{I. Shavitt,J. Hausinger}, in the basis of \{\(\ket{1,-\kappa\pm},\ket{0,0-},\ket{1,\kappa\pm}\)\} (using this ordering), the first-order resonance dynamics of the system can be described by an effective three-level model (for more details, see Appendix \ref{Appendix B}),
	\begin{equation}\label{eq16}
		T_{\pm}(\kappa)\simeq \begin{pmatrix}
			E_-(0)& \Gamma_{\pm,1}(\kappa) &0\\
			\Gamma_{\pm,1}(\kappa)&E_-(0) &\Gamma_{\pm,2}(\kappa)\\
			0& \Gamma_{\pm,2}(\kappa)&E_-(0)
		\end{pmatrix},
	\end{equation}
	where 
	\begin{align}\label{eq17}
		&\Gamma_{-,1}(\kappa)=-\frac{K}{4}\cos[\theta(-\kappa)-\theta(0)],\\\nonumber &\Gamma_{-,2}(\kappa)=\frac{K}{4}\cos[\theta(\kappa)-\theta(0)],\\\nonumber &\Gamma_{+,1}(\kappa)=-\frac{K}{4}\sin[\theta(-\kappa)-\theta(0)],\\\nonumber
		&\Gamma_{+,2}(\kappa)=\frac{K}{4}\sin[\theta(\kappa)-\theta(0)].\\\nonumber
	\end{align}
	The “-” in the plus-minus sign in Eq. (\ref{eq16}) signifies the resonance between the ground state and the excited states in the lower energy band, whereas the “+” sign denotes the resonance between the ground state and the excited states in the upper energy band, as schematically depicted in Figs. \ref{fig2} (b) and (c), respectively. The off-diagonal matrix elements $\Gamma_{\pm,1}$ ($\Gamma_{\pm,2}$) correspond to the effective coupling strength between the initial state and the excited states. The index 1 in $\Gamma_{\pm,1}$ represents the coupling between the ground state and the excited states \(\ket{-\kappa\pm}\), whereas index 2 denotes the coupling between the ground state and the excited states \(\ket{\kappa\pm}\) that are· degenerate with \(\ket{-\kappa\pm}\).
	From a symmetry perspective, it is observed that \(\sin[\theta(-n)] = \cos[\theta(n)]\) and \(\cos[\theta(-n)] = \sin[\theta(n)]\), resulting in \( |\Gamma_{\pm,1}(\kappa)| = |\Gamma_{\pm,2}(\kappa)| \). By introducing a rotation \(\{\ket{1,-\kappa\pm},\ket{0,0-},\ket{1,\kappa\pm}\}\) \(\rightarrow\) \(\{\ket{a(\kappa)\pm},\ket{0,0-},\ket{b(\kappa)\pm}\}\), 
	\begin{align}\label{eq18}
	&\ket{a(\kappa)\pm}=\frac{1}{\sqrt{\Gamma^2_{\pm,1}+\Gamma^2_{\pm,2}}}(\Gamma_{\pm,1}\ket{1,-\kappa\pm}+\Gamma_{\pm,2}\ket{1,\kappa\pm}),\\\nonumber
	&\ket{b(\kappa)\pm}=\frac{1}{\sqrt{\Gamma^2_{\pm,1}+\Gamma^2_{\pm,2}}}(\Gamma_{\pm,1}\ket{1,-\kappa\pm}-\Gamma_{\pm,2}\ket{1,\kappa\pm}),
	\end{align}
	we find that the basis state \(\ket{b(\kappa)\pm}\) is decoupled from \(\ket{0,0-}\), and in the rotated orthonormal basis, Eq. (\ref{eq16}) is transformed to
	\begin{equation}\label{eq19}
		\tilde{T}_{\pm}(\kappa)\simeq \begin{pmatrix}
			E_-(0)& \Gamma_{\pm}(\kappa) &0\\
			\Gamma_{\pm}(\kappa)&E_-(0) &0\\
			0& 0&E_-(0)
		\end{pmatrix},
	\end{equation}
	where \(\Gamma_{\pm}(\kappa) = \bra{a(\kappa)\pm}T_{\pm}(\kappa)\ket{0,0-} = \sqrt{\Gamma^2_{\pm,1} + \Gamma^2_{\pm,2}}\), yielding
	\begin{align}\label{eq20}
	&\Gamma_{+}(\kappa)=\frac{K}{4}\sqrt{\sin^2[\theta(-\kappa)-\theta(0)]+\sin^2[\theta(\kappa)-\theta(0)]}, \\\nonumber
	&\Gamma_{-}(\kappa)=\frac{K}{4}\sqrt{\cos^2[\theta(-\kappa)-\theta(0)]+\cos^2[\theta(\kappa)-\theta(0)]},
    \end{align}
	which represents the coupling strength between the initial state \(\ket{0,0-}\) and the state \(\ket{a(\kappa)\pm}\). In this case, the Rabi oscillation  occurs between the initial state \(\ket{0-}\) and state $\ket{a(\kappa)\pm}$, and the analytical prediction for time-averaged canonical momentum of different spin components ($\sigma=\uparrow$ or $\downarrow$) is given by
	\begin{equation}\label{eq21}
	\bar{p}^\pm_{\sigma}(\kappa)=\frac{p^{a\pm}_{\sigma}(\kappa)}{2},
	\end{equation}
	where \( p^{a\pm}_{\sigma}(\kappa) = \int_0^{2\pi}dx\big[\psi^{a\pm}_\sigma(x)\big]^*\hat{p}\psi^{a\pm}_\sigma(x) \) represents the canonical momentum of the different components of the SO-coupled BEC carried by \(\ket{a(\kappa)\pm}=[\psi^{a\pm}_\uparrow(x), \psi^{a\pm}_\downarrow(x)]^T\), which is given by 
	\begin{align}\label{eq22}
	&p^{a-}_{\uparrow}(\kappa)=\frac{\kappa}{\Gamma^2_-(\kappa)}\bigg(-\Gamma^2_{-,1}(\kappa)\cos^{2}\theta(-\kappa)+\Gamma^2_{-,2}(\kappa)\cos^{2}\theta(\kappa)\bigg),\\\nonumber
	&p^{a-}_{\downarrow}(\kappa)=\frac{\kappa}{\Gamma^2_-(\kappa)}\bigg(-\Gamma^2_{-,1}(\kappa)\sin^{2}\theta(-\kappa)+\Gamma^2_{-,2}(\kappa)\sin^{2}\theta(\kappa)\bigg),\\\nonumber
	&p^{a+}_{\uparrow}(\kappa)=\frac{\kappa}{\Gamma^2_+(\kappa)}\bigg(-\Gamma^2_{+,1}(\kappa)\sin^{2}\theta(-\kappa)+\Gamma^2_{+,2}(\kappa)\sin^{2}\theta(\kappa)\bigg),\\\nonumber
	&p^{a+}_{\downarrow}(\kappa)=\frac{\kappa}{\Gamma^2_+(\kappa)}\bigg(-\Gamma^2_{+,1}(\kappa)\cos^{2}\theta(-\kappa)+-\Gamma^2_{+,2}(\kappa)\cos^{2}\theta(\kappa)\bigg).\nonumber
	\end{align}
	As before, the plus-minus sign in Eq. (\ref{eq21}) denotes the resonance between the ground state \(\ket{0-}\) and the excited states in the upper and lower energy bands, respectively.
	
	Interestingly, regardless of whether the $\pm$ sign is positive or negative, the signs of the canonical momenta for the up and down components are always opposite and equal in magnitude, as expressed by \(\bar{p}^-_{\uparrow}(\kappa) = -\bar{p}^-_{\downarrow}(\kappa) > 0\) and \(\bar{p}^+_{\downarrow}(\kappa) = -\bar{p}^+_{\uparrow}(\kappa) > 0\), which provides a means to control the canonical momentum of the up and down components in a SO-coupled BEC. We can also grasp this mechanism from a different perspective. The Hamiltonian, as defined by Eq. (\ref{eq5}), is symmetric under the combined action of the spin-flip operator \(\hat{\sigma}_x\) and the time-inversion operator \(\hat{\mathcal{T}}: i \to -i, \hat{p} \to -\hat{p}\), leading to \(\hat{\sigma}_x \hat{\mathcal{T}} \ket{-n\pm} = \pm\ket{n\pm}\).  This implies that: when compared to the eigenstates \(\ket{n\pm}\), the eigenstates \(\ket{-n\pm}\) carry negative canonical momentum, and the populations of the spin components are inverted. Given the equal coupling strengths \(|\Gamma_{\pm,1}(\kappa)| = |\Gamma_{\pm,2}(\kappa)|\), the state \(\ket{a(\kappa)\pm}\) as described by Eq. (\ref{eq18}) carries pure canonical spin  currents, characterized by \(p^{a\pm}_{\uparrow}(\kappa) - p^{a\pm}_{\downarrow}(\kappa) \neq 0\) and \(p^{a\pm}_{\uparrow}(\kappa) + p^{a\pm}_{\downarrow}(\kappa) = 0\). From a symmetry perspective, as the Rabi oscillation occurs between the initial state \(\ket{0-}\) and the state \(\ket{a(\kappa)\pm}\), the population imbalance between the two components consistently remains at zero, that is, $\Delta N(t)=0$. This implies that the system maintains a perfectly symmetric state where the number of particles with up-spin and down-spin are equal. In such a scenario, only a pure alternating spin current is generated, without any accompanying mass current.
	According to the relationship between the spin (mass) current and the canonical momentum, as expressed in Eqs. (\ref{eq12}) and (\ref{eq13}), the time-averaged spin (mass) current is analytically determined by
	\begin{equation}\label{eq23}
		\bar{I}^\pm_{s}(\kappa)=\bar{p}^\pm_{\uparrow}(\kappa)-\bar{p}^\pm_{\downarrow}(\kappa)-k_0,\quad\bar{I}^\pm_{m}(\kappa)=0.
	\end{equation}
	\begin{figure}[htp]	
		\centering
		\includegraphics[width=9cm]{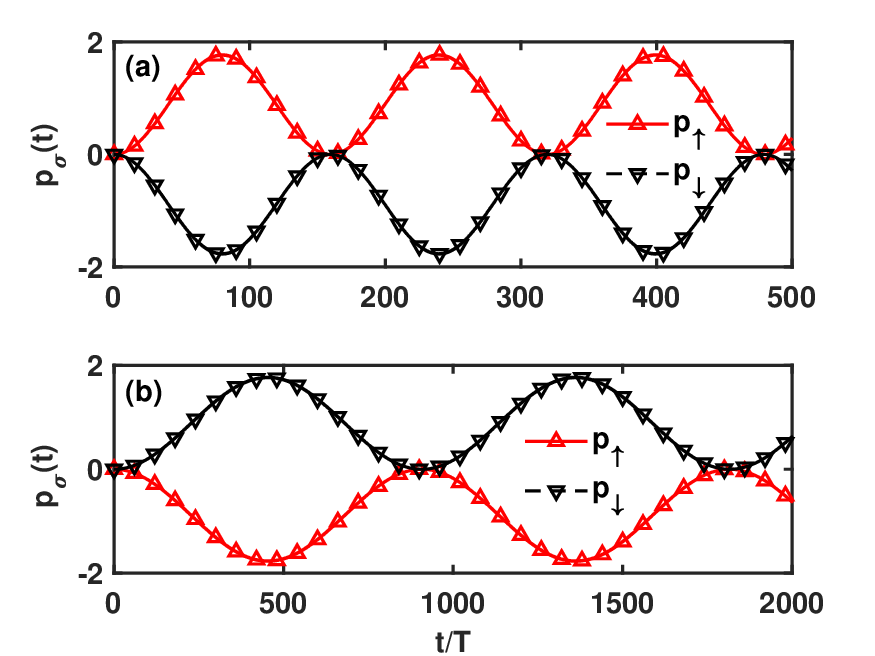}
		\caption{ Comparison of the exact numerical results obtained by Eq. (\ref{eq1}) (shown by lines) with the analytical results predicted by the effective three-level model $T_\pm(\kappa)$ in Eq. (\ref{eq16}) (triangles) with weak driving $K=0.1$. The solid (dashed) line and up (down) triangles mark the canonical momentum $p_\uparrow$ ($p_\downarrow$) for the spin components of \(\uparrow\) (\(\downarrow\)). (a) Time evolution of $p_\uparrow$ ($p_\downarrow$)  for the resonant coupling of the initial ground state to the excited states in the lower branch \( E_-(\kappa) \), as schematically shown in Fig. \ref{fig2} (b). (b) Time evolution of $p_\uparrow$ ($p_\downarrow$) for the resonant coupling of the initial ground state to the excited states in the upper branch \( E_+(\kappa) \), as schematically shown in Fig. \ref{fig2} (c). The other parameters are $\kappa=5$, $k_0=1$, $\Omega=10$, $\omega=E_-(\kappa)-E_-(0)$ for (a) and $\omega=E_+(\kappa)-E_-(0)$ for (b).} \label{fig3}
	\end{figure}
	\begin{figure}[htp]	
		\centering
		\includegraphics[width=9cm]{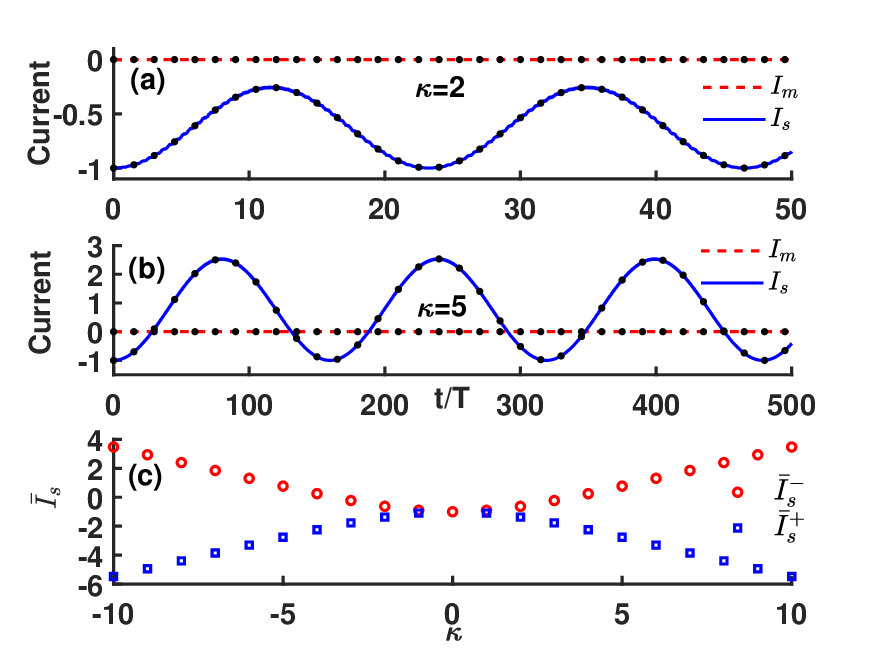}
		\caption{(a): Pure alternating (AC) spin current with zero mass current induced by the first-photon resonance at frequency \( \omega = E_-(\kappa) - E_-(0) \) with \( \kappa = 2 \). (b): Pure alternating (AC) spin current with zero mass current induced by the first-photon resonance at frequency \( \omega = E_-(\kappa) - E_-(0) \) with \( \kappa = 5 \). (c): Time-averaged spin current $\bar{I}_s^\pm$ predicted by Eqs. (\ref{eq12}) and (\ref{eq21}) versus \(\kappa\). Here  circles are  for $\bar{I}_s^-$ induced by resonant coupling of the initial ground state to lower branch \( E_-(n) \),  and   squares are for $\bar{I}_s^+$ induced by resonant coupling of the initial ground state to upper branch \( E_+(n) \). The other parameters are \(K=0.1\), \(k_0=1\), \(\Omega=10\). In (a) and (b), the exact numerical results obtained by Eq. (\ref{eq1}) are depicted by lines, while the analytical results predicted by the effective three-level model (\ref{eq16}) are illustrated by dots.} \label{fig4}
	\end{figure}
	To validate our theoretical approach for generating a purely alternating (AC) spin current devoid of mass current through photon-assisted tunneling (resonance), we present numerical results for the time evolution of the canonical momentum for each spin component, obtained by integrating Eq. (\ref{eq1}) using the split-step Fourier (SSF) method with an initial state $\ket{0-}$,  as shown in Fig. \ref{fig3}. These results are then compared with the analytical predictions from Eq. (\ref{eq19}), revealing an exceptional degree of agreement. The first-photon resonance dynamics, which involve the mixing of  the initial ground state $\ket{0-}$ with the excited states $\ket{\pm\kappa-}$ and $\ket{\pm\kappa+}$ in lower branch \(E_-(\kappa)\) and upper branch \(E_+(\kappa)\), respectively, for a fixed wave number of the spatial potential \( \kappa = 5 \), are illustrated in Figs. \ref{fig3} (a) and \ref{fig3} (b), where the time evolution of the canonical momentum for each component exhibits Rabi oscillation with identical amplitude and frequency but in opposite directions. By comparing Figs. \ref{fig3} (a) and \ref{fig3} (b), as expected, we note that the resonance dynamics of the ground state coupling to the lower and upper energy branches result in distinct flow patterns for the two spin components, with the flow directions being reversed in the two scenarios. By utilizing the symmetry properties, we are aware that the population difference between the two components remains zero throughout the evolution. As a result, we are able to generate solely an AC spin current without the presence of a mass current, which has been numerically validated in Figs. \ref{fig4} (a) and \ref{fig4} (b) with different wave vectors $\kappa=2$ and $\kappa=5$. The results show that the time-averaged spin current increases with the wave vector \(\kappa\), and there is no mass current, as depicted in Fig. \ref{fig4} (c). This is reasonable, as the frequency \(\omega\) provides the energy necessary to mix the ground state with the excited states, and similarly, the wave vector \(\kappa\) provides the momentum required to link these two states. Consequently, as the wave vector \(\kappa\) increases, the resulting spin current also increases. The results indicate that photon-assisted resonance provides a straightforward and effective method for generating and controlling the net spin current, with both its magnitude and direction being easily and precisely controlled.

	\section{Double well phase} \label{V}
\begin{figure}[htp]  
		\centering
		\includegraphics[width=9cm]{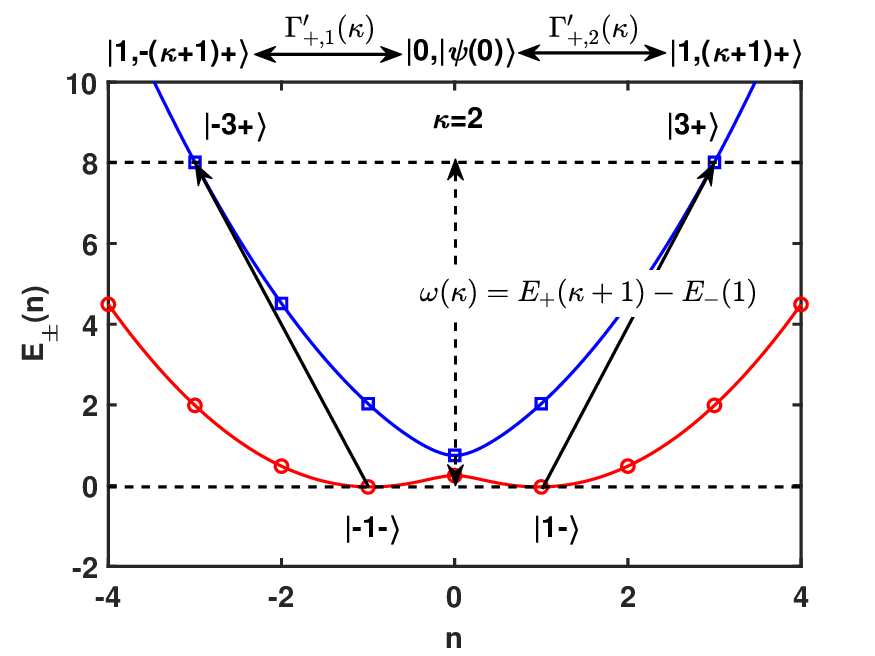}
		\caption{Schematic representation of the first photon resonance in the double-well phase: the initial state \(\ket{\psi(0)}=c_{-}\ket{-1-}+c_+\ket{1-}\) is coupled to the two degenerate excited states \(\ket{\pm(\kappa+1)}\) via the resonant driving with the resonance frequency \(\omega(\kappa) = E_+(\kappa+1) - E_-(1)\). The effective coupling strengths are \(\Gamma_{+,1}^\prime(\kappa)\) and \(\Gamma_{+,2}^\prime(\kappa)\). Here, \(\kappa = 2\) is used as an example for illustration purposes.} \label{fig5}
	\end{figure}

	In this section, we will delve into mechanisms for the generation and control of both spin and mass currents by photon-assisted resonance in the double-well phase, where the single-particle dispersion exhibits two degenerate minima, as described in Eq. (\ref{eq9}). In this instance, the initial state is prepared as a linear superposition of the degenerate states \(\ket{-1-}\) and \(\ket{1-}\),
	\begin{equation}\label{eq24}
		\ket{\psi(0)}=c_{-}\ket{-1-}+c_+\ket{1-},
	\end{equation} 
	where \( c_- \) and \( c_+ \) are the projection coefficients of the initial state onto the degenerate states \(\ket{-1-}\) and \(\ket{1-}\), respectively, such that \( c_{\pm} = \braket{\pm1-|\psi(0)} \). Here, if we choose the Raman coupling strength $\Omega$ to be $E_r$ or less, $\bar{n}$ (in units of $k_0$), as defined in Eq. (\ref{eq9}), is close to $n = ±1$, and the initial state (\ref{eq24}) is nearly aligned with the ground state. As discussed earlier, the resonant driving frequency bridges the energy gap between the initial state and other resonant states, and the canonical momentum exchange between the ground state and the excited states is provided by the wave vector $\kappa$. What we can expect is that there are two ways to exchange canonical momentum: Case I involves the state with positive momentum, \(\ket{1-}\), coupling to a state with positive momentum, \(\ket{(\kappa+1)+}\) (for instance, we consider only the coupling to the \(E_+(n)\) branch), in the positive direction, and the state with negative momentum, \(\ket{-1-}\), coupling to a state with negative momentum, \(\ket{-(\kappa+1)+}\), in the negative direction. Case II describes a scenario where the state \(\ket{1-}\) with positive momentum couples to a state with negative momentum, such as \(\ket{(-\kappa+1)+}\), in the negative direction, and the state \(\ket{-1-}\) with negative momentum couples to a state with positive momentum, such as \(\ket{(\kappa-1)+}\), in the positive direction. The illustration for Case I is presented in Fig. \ref{fig5}, which is the specific scenario that we are considering in this paper. By following the same derivation process as for Eq. (\ref{eq16}), we treat \(H_0\) as the unperturbed term and consider the weak driving \(V(x,t)\) as a perturbation within the extended Hilbert space. Given the initial state as defined in Eq. (\ref{eq24}), we can anticipate that the three resonant states \(\ket{1,-(\kappa+1)+},\ket{0,\psi(0)},\ket{1,(\kappa+1)+}\) within the extended Hilbert space will be linked by the flashing potential at the first photon resonance frequency \(\omega = E_+(\kappa+1) - E_-(1)\). Therefore, a truncated Hilbert space spanned solely by the Floquet states \(\ket{1,-(\kappa+1)+},\ket{0,\psi(0)},\ket{1,(\kappa+1)+}\) may suffice to describe the dynamics under weak driving, and the corresponding dynamics for first photon resonance can be described by an effective three-level model (for more details, see Appendix \ref{Appendix B}),
	\begin{equation}\label{eq25}
		T_{+}(\kappa)\simeq \begin{pmatrix}
			E_-(1)& \Gamma^\prime_{+,1}(\kappa) &0\\
			\Gamma^{\prime*}_{+,1}(\kappa)&E_-(1) &\Gamma^{\prime*}_{+,2}(\kappa)\\
			0& \Gamma^\prime_{+,2}(\kappa)&E_-(1)
		\end{pmatrix},
	\end{equation}
	where \( * \) denotes the complex conjugation and the effective coupling is given by 
	\begin{align}\label{eq26}
		&\Gamma^\prime_{+,1}(\kappa)=-\frac{Kc_-}{4}\sin[\theta(-\kappa-1)-\theta(-1)],\\\nonumber
		&\Gamma^\prime_{+,2}(\kappa)=\frac{Kc_+}{4}\sin[\theta(\kappa+1)-\theta(1)].\\\nonumber
	\end{align} 
	Following the same procedure as mentioned in the case of the single-well phase, we introduce the new orthonormal state basis \(\{\ket{a^\prime(\kappa)+},\ket{0,\psi(0)},\ket{b^\prime(\kappa)+}\}\) for the matrix $T_+(\kappa)$ in Eq. (\ref{eq25}), with 
	\begin{align}\label{eq27}
		&\ket{a^\prime(\kappa)+}=\frac{1}{\Gamma^\prime_{+}(\kappa)}[\Gamma^{\prime}_{+,1}\ket{1,-(\kappa+1)+}+\Gamma^{\prime}_{+,2}\ket{1,(\kappa+1)+}],\\\nonumber
		&\ket{b^\prime(\kappa)+}=\frac{1}{\Gamma^\prime_{+}(\kappa)}[\Gamma^{\prime}_{+,1}\ket{1,-(\kappa+1)+}-\Gamma^{\prime}_{+,2}\ket{1,(\kappa+1)+}],
	\end{align} 
	and we find that $\bra{b^\prime(\kappa)+}T_+(\kappa)\ket{0,\psi(0)} =\bra{a^\prime(\kappa)+}T_{+}\ket{b^\prime(\kappa)+}= 0$ and 
	\(\Gamma^\prime_{+}(\kappa) = \bra{a^\prime(\kappa)+}T_{+}(\kappa)\ket{0,\psi(0)} = \sqrt{|\Gamma^{\prime}_{+,1}|^2 + |\Gamma^{\prime}_{+,2}|^2}\). Therefore, the three-level model $T_+(\kappa)$ in Eq. (\ref{eq25}) can be transformed to an effective two-level problem, in which Rabi oscillation between $\ket{a^\prime(\kappa)+}$ and $\ket{0,\psi(0)}$ occurs with the oscillation period $\pi/\Gamma^\prime_{+}(\kappa)$, and the system, on average, spends  equal amounts of time in the state $\ket{a^\prime(\kappa)+}$ and $\ket{0,\psi(0)}$. The analytical prediction for the time-averaged canonical momentum of different spin components ($ \sigma=\uparrow$ or $\downarrow$)  is then simply expressed as
	\begin{equation}\label{eq28}
		\bar{p}^+_{\sigma}(\kappa)=\frac{p^{\psi(0)}_{\sigma}+p^{a^\prime+}_{\sigma}(\kappa)}{2},
	\end{equation}
	where $p^{\psi(0)}_{\sigma}$ and $p^{a^\prime+}_{\sigma}(\kappa)$ are the canonical momentum of different components carried by the initial state $\ket{\psi(0)}$ and the state $\ket{a^\prime(\kappa)+}$ respectively, with
	\begin{align}\label{eq29}
		p^{\psi(0)}_\uparrow&=-|c_-|^2\cos^2\theta(-1)+|c_+|^2\cos^2\theta(1),\\\nonumber
		p^{\psi(0)}_\downarrow&=-|c_-|^2\sin^2\theta(-1)+|c_+|^2\sin^2\theta(1),\\\nonumber
		p^{a^\prime+}_{\uparrow}(\kappa)&=\frac{\kappa+1}{\Gamma^{\prime2}_+(\kappa)}\bigg(-\Gamma^{\prime2}_{+,1}(\kappa)\sin^{2}\theta[-(\kappa+1)]\\\nonumber
		&\quad+\Gamma^{\prime2}_{+,2}(\kappa)\sin^{2}\theta(\kappa+1)\bigg),\\\nonumber
		p^{a^\prime+}_{\downarrow}(\kappa)&=\frac{\kappa+1}{\Gamma^{\prime2}_+(\kappa)}\bigg(-\Gamma^{\prime2}_{+,1}(\kappa)\cos^{2}\theta[-(\kappa+1)]\\\nonumber
		&\quad +\Gamma^{\prime2}_{+,2}(\kappa)\cos^{2}\theta(\kappa+1)\bigg).
	\end{align}
	\begin{figure}[htp]	
		\centering
		\includegraphics[width=9cm]{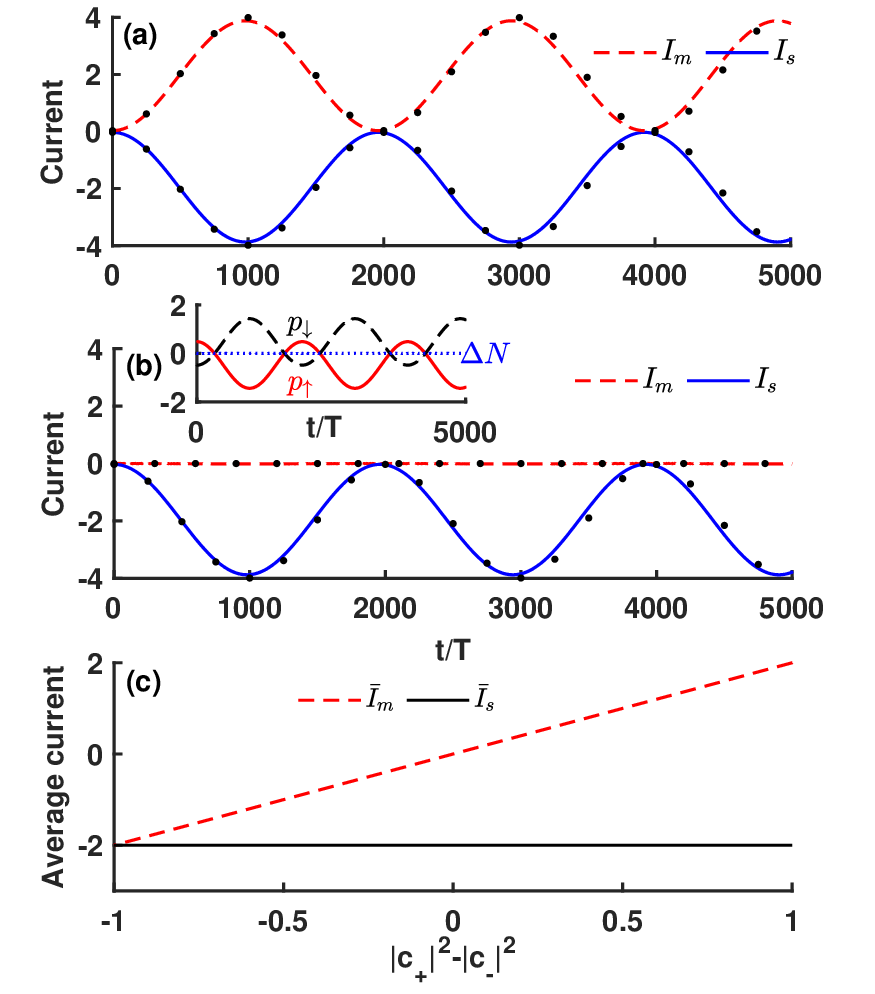}
		\caption{(a) and (b): Comparison between the exact numerical results obtained from Eq. (\ref{eq1}) (lines) with the analytical results predicted by the effective three-level model $T_+(\kappa)$ (dots) in Eq. (\ref{eq25}) under weak driving conditions, with \( K = 0.1 \). The solid line represents the time evolution of the spin current, while the dashed line indicates the time evolution of the mass current. (a) The initial state is prepared according to Eq. (\ref{eq24}) with \( c_- = 0 \) and \( c_+ = 1 \); (b) the initial state is set with \( c_- = c_+ = 1/\sqrt{2} \). (c) The time-averaged spin current (solid line) and mass current (dashed line) are plotted against the population imbalance \( |C_2|^2 - |C_1|^2 \). The parameters are \( \kappa = 2, k_0 = 1, \Omega = 0.5, \) and \( \omega = E_+(1+\kappa) - E_-(0) \). In panel (b), an inset is provided to clearly illustrate the time evolution of $p_\uparrow(t)$ (solid line) and $p_\downarrow(t)$ (dashed line), and we also numerically witness that $\Delta N(t)$ equals zero (indicated by the blue dotted line) for the initial state Eq. (\ref{eq24}), where $c_-$ equals $c_+$, both set to $1/\sqrt{2}$.} \label{fig6}
	\end{figure}
	Fig. \ref{fig6} shows a very good agreement between the numerical results obtained by Eq. (\ref{eq1}) and the analytical results predicted by Eq. (\ref{eq25}) for the time-dependent spin and mass currents. Here, we consider two distinct initial states: one is an unequally weighted superposition, with \( |c_-|^2 = 0 \) and \( |c_+|^2 = 1 \), between the modes \(\ket{-1-}\) and \(\ket{1-}\), as illustrated in Fig. \ref{fig6} (a); the other is an equally weighted superposition, with \( |c_-|^2 = |c_+|^2 = \frac{1}{2} \), between the modes \(\ket{-1-}\) and \(\ket{1-}\), as illustrated in Fig. \ref{fig6} (b). For the former, we observe that both AC mass current and AC spin current are generated. For the latter, a pure spin current is generated with a vanishing mass current, as the mass current of the initial state and the state \(\ket{a^\prime(\kappa)+}\) both have a zero value. However, in both of these two conditions, the time evolution of the spin current exhibits identical Rabi oscillations. Furthermore, we investigate the time-averaged spin and mass currents versus the probability imbalance \( |c_+|^2 - |c_-|^2 \) in Fig. \ref{fig6} (c). It is clearly seen that the time-averaged mass current is linearly dependent on \( |c_+|^2 - |c_-|^2 \), while the time-averaged spin current is independent of \( |c_+|^2 - |c_-|^2 \).   In fact, this mechanism can be analytically derived: according to Eqs. (\ref{eq28}) and (\ref{eq29}) as well as Eqs. (\ref{eq12}) and (\ref{eq13}), the time-averaged spin current is found to be independent of the probability imbalance \( |c_+|^2 - |c_-|^2 \), as evidenced by 
	\begin{align}\label{eq30}
	\bar{I}_s(\kappa)&=\bar{p}^+_{\uparrow}(\kappa)-\bar{p}^+_{\downarrow}(\kappa)-k_0\\\nonumber
	&=\frac{-(\kappa+1)\cos[2\theta(\kappa+1)]+\cos2[\theta(1)]}{2}-k_0,
	\end{align}
	and the time-averaged mass current is linearly dependent on the probability imbalance \( |c_+|^2 - |c_-|^2 \), which is given by
	\begin{align}\label{eq31}
	\bar{I}_m(\kappa)&=\bar{p}^+_{\uparrow}(\kappa)+\bar{p}^+_{\downarrow}(\kappa)-k_0\overline{\Delta N}\\\nonumber
	&=\frac{(\kappa+2)\bigg(|c_+|^2-|c_-|^2\bigg)}{2}.
	\end{align}
	In the derivation of Eq. (\ref{eq31}), we have utilized the result that the time-averaged population imbalance is zero, i.e., \(\overline{\Delta N} = 0\).

	\section{Conclusion} \label{VI}
	In summary, we present a scheme based on the photon-assisted tunneling (resonance) to generate and control the spin and mass current in noninteracting BEC (or a SO-coupled atom) in a toroidal trap using a flashing potential. By modulating the Raman coupling strength, the ground band exhibits two distinct phases, leading to markedly different dynamic behaviors of the spin and mass currents in each phase. When the single-particle dispersion assumes a single-well structure, a pure alternating (AC) spin current without mass current is realized through a single-photon resonance scheme. When the dispersion features two degenerate minima, the spin current dynamics remain unchanged for any choice of the superposition of degenerate single-particle ground states, with the time-averaged mass current linearly dependent on the imbalance in the occupation probabilities between the two degenerate ground states. In both phases, by changing the driving frequency and the wave vector of the flashing potential, which provide photon-assisted energy matching and momentum transfer between the initial ground state and the resonant excited state, respectively, we are able to achieve controllable alternating (AC) spin and mass currents with tunable direction and magnitude. The use of a flashing potential provides a means to manipulate the spin and momentum of the SO-coupled BEC. By carefully controlling the driving frequency and strength of the flashing potential, we can induce specific spin dynamics and observe their consequences. We note that BECs in toroidal traps have been successfully realized in various experiments \cite{S. Gupta,L. Corman,G. Del Pace}, which lays a foundation for exploring spin transport in SO-coupled BECs using a flashing ring-shaped potential.
	\acknowledgments
	The work was supported by  the National
	Natural Science Foundation of China (Grant Nos. 12375022, 11975110), the Natural Science Foundation of Zhejiang Province (Grant No. LY21A050002), and Zhejiang Sci-Tech University Scientific Research
	Start-up Fund (Grant No. 20062318-Y).

	\appendix
	\section{The definition of spin current and mass current}\label{Appendix A}
	We start with the Hamiltonian for a Raman-induced SO-coupled noninteracting BEC driven by the flashing potential 
	\begin{equation}\label{A.1}
		\hat{H}=\frac{\hat{p}^2}{2}-k_0\hat{p}\hat{\sigma}_z+\frac{k_0^2}{2}+\frac{\Omega}{2}\hat{\sigma}_x+\hat{V}(\hat{x},t).\tag{A.1}
	\end{equation}
	The spin current is defined as the difference in the average flow of particles between the spin-up and spin-down components. The spin-current operator along the $x$-direction can be defined as \(\hat{J}^s_x=\hat{x}\hat{\sigma_z}\), and the Heisenberg equation of motion for this operator is as follows:
	\begin{align}\label{A.2}
		\frac{\mathrm{d}}{\mathrm{d}t}\hat{J}^s_x = \frac{1}{i}\left[\hat{J}^s_x, \hat{H}\right]=\Omega \hat{x}\hat{\sigma}_y+\hat{p}\hat{\sigma}_z-k_0.\tag{A.2}
	\end{align}
	When calculating the spin current, the term \(\Omega\hat{x}\hat{\sigma}_y\) in Eq. (\ref{A.2}) does not contribute to the spin current calculation, due to the fact that the expectation value \(\braket{\hat{\sigma}_y}\) is zero, stemming from the exclusion of the Pauli operator \(\hat{\sigma}_y\) from the Hamiltonian given in Eq. (\ref{A.1}). Therefore, the spin current is given by
	\begin{equation}\label{A.3}
		I_s=\frac{\mathrm{d}}{\mathrm{d}t}\braket{\hat{J}^s_x}=\braket{\hat{p}\hat{\sigma}_z}-k_0=p_\uparrow-p_\downarrow-k_0,\tag{A.3}
	\end{equation}
	which defines the spin current in terms of the difference of mean canonical momentum shifted by the laser wave vector.

	The mass current is defined as the expectation value of the velocity operator. The velocity operator corresponding to the mechanical momentum  is given by
	\begin{equation}\label{A.4}
		\frac{\mathrm{d}\hat{x}}{\mathrm{d}t}=\frac{1}{i}\big[\hat{x},\hat{H}]=\hat{p}-k_0\hat{\sigma}_z.\tag{A.4}
	\end{equation}
	Therefore, the mass current is defined as the mean canonical momentum minus \( k_0 \) times the imbalance of the population imbalance between the two components, which is explicitly expressed as
	\begin{equation}\label{A.5}
		I_m=	\frac{\mathrm{d}\braket{\hat{x}}}{\mathrm{d}t}=\braket{\hat{p}}-k_0\braket{\hat{\sigma}_z}=p_\uparrow+p_\downarrow-k_0\Delta N.\tag{A.5}
	\end{equation}

	\section{Derivation of effective three-level model}\label{Appendix B}
	Under the resonance condition \( E_\pm(\kappa) - E_g = m\omega \), the unperturbed states \(\{\ket{1,-\kappa\pm},\ket{0,0-},\ket{1,\kappa\pm}\}\) become degenerate. We anticipate that the weak driving will mix these three resonant states. According to the degenerate perturbative method in the extended Hilbert space, the effective dynamics up to the first-order can be described by an effective \(T\) matrix with the matrix elements
	\begin{equation}  \label{C.1}
		\bra{m,n\alpha}\hat{T}\ket{m^\prime,n^\prime\alpha^\prime}\simeq \bra{m,n\alpha}\Big(H_0+\hat{V}\Big)\ket{m^\prime,n^\prime\alpha^\prime}.\tag{B.1}
	\end{equation}
	In our model, the driving reads
	\begin{align}\nonumber  \label{C.2}
		\hat{V}&=\frac{K}{4}(e^{-i\omega t}-e^{i\omega t})(e^{i\kappa x}-e^{-i\kappa x})\\
		&=\frac{K}{4}\sum_{m}(\ket{m}\bra{m-1}-\ket{m}\bra{m+1})\otimes(e^{i\kappa x}-e^{-i\kappa x}).
		\tag{B.2}
	\end{align}
	For example, the off-diagonal terms of matrix $T_+$ in Eq.  (\ref{eq16}) can be calculated as follows:
	\begin{align} \nonumber \label{C.3}
	\Gamma_{+,1}(\kappa)&\simeq \bra{1,-\kappa+}\hat{V}\ket{0,0-}\\\nonumber
	&=-\frac{K}{4}\bra{-\kappa+}e^{-i\kappa x}\ket{0-}\\
	&=-\frac{K}{4}		\begin{pmatrix}
		\sin\theta(-\kappa) & \cos\theta(-\kappa)
	\end{pmatrix}\cdot
	\begin{pmatrix}
	\cos\theta(0) \\ -\sin\theta(0)\nonumber
	\end{pmatrix}\\
	&=-\frac{K}{4}\sin\big[\theta(-\kappa)-\theta(0)],
	\tag{B.3}
	\end{align}
	and
	\begin{align} \nonumber \label{C.4}
	\Gamma_{+,2}(\kappa)&\simeq \bra{1,\kappa+}\hat{V}\ket{0,0-}\\\nonumber
	&=\frac{K}{4}\bra{\kappa+}e^{-i\kappa x}\ket{0-}\\
	&=\frac{K}{4}		\begin{pmatrix}
		\sin\theta(\kappa) & \cos\theta(\kappa)
	\end{pmatrix}\cdot
	\begin{pmatrix}
		\cos\theta(0) \\ -\sin\theta(0)\nonumber
	\end{pmatrix}\\
	&=\frac{K}{4}\sin\big[\theta(\kappa)-\theta(0)]
	\tag{B.4}.
	\end{align}
	The diagonal terms of matrix $T_+$ in Eq. (\ref{eq16}) are given by \(\bra{1,-\kappa+}(H_0-\omega)\ket{1,-\kappa+}=\bra{0,0-}H_0\ket{0,0-} =\bra{1,\kappa+}(H_0-\omega)\ket{1,\kappa+}= E_-(0)\). Here, we only present the calculation process for the \(T_+\) matrix involving the mixing of the ground state \(\ket{0,0-}\) to the excited states $\ket{-\kappa+}$ and $\ket{\kappa+}$ in the upper branch $E_+(n)$. The calculation for the matrix $T_-$ in Eq. (\ref{eq16}) can be obtained in a similar way. Following the same procedure as above, under the resonance condition \(\omega = E_+(\kappa+1) - E_-(1)\), we can construct the effective three-level model for the double-well phase, with the matrix defined by Eq. (\ref{eq25}) in the main text, which only involves the first-order transition between the resonant Floquet states \(\ket{1,-(\kappa+1)+},\ket{0,\psi(0)},\ket{1,(\kappa+1)+}\).

\end{document}